\documentclass[prd,preprint,aps,amsfonts,amssymb,nofootinbib,tightenlines]{revtex4}
\usepackage{amscd}
\usepackage{amsmath}
\usepackage{bm}
\usepackage[dvips]{color}                 
\usepackage{graphicx}
\usepackage{eufrak}
\usepackage{eucal}

\newcommand{\beq}{\begin{equation}}
\newcommand{\eeq}{\end{equation}}
\newcommand{\bea}{\begin{eqnarray}}
\newcommand{\eea}{\end{eqnarray}}

\newcommand{\benn}{\begin{displaymath}}
\newcommand{\eenn}{\end{displaymath}}

\def\W{{\cal W}}

\begin{document}

\title{\bf  Phase Transition in Asymmetrical Superfluids I: \\Equal Fermi Surfaces}

\author{Heron Caldas\footnote{{\tt hcaldas@ufsj.edu.br}}, C.W. Morais and A.L. Mota\footnote{{\tt motaal@ufsj.edu.br}}}
\affiliation{Universidade Federal de S\~{a}o Jo\~{a}o del Rey, S\~{a}o Jo\~{a}o del Rei, 36300-000, MG, Brazil}

\begin{abstract}
In this paper, we study phase transitions in asymmetrical fermion superfluids. In this scenario, the candidates to form pair are particles with mismatched masses and chemical potentials. We derive an expression for the critical temperature in terms of the gap and masses (or chemical potentials) when the constraint of equal Fermi surfaces $m_a\mu_a = m_b\mu_b$ is imposed.
\end{abstract}
\maketitle
\bigskip

The advent of new techniques to deal with ultracold fermionic atoms has motivated the interest on its theoretical investigation. One of the most intriguing systems that have been investigated in last few years are the asymmetric ones. In these asymmetrical Fermi systems the masses and densities or chemical potentials of the two species that will possibly form pair are unequal. Candidates for the ground states of these cold, dilute quantum systems have been proposed: the Sarma phase~\cite{Sarma}, the Larkin-Ovchinnikov-Fulde-Ferrel (LOFF) phase~\cite{loff}, the breached-pair superfluidity (BP) phase~\cite{Wilczek1,Wilczek2,Wilczek3}, and the mixed phase (MP)~\cite{Heron1,Heron2}. Pairing in asymmetrical fermionic systems are predict to have applications from the explanation of chirality among amino acids~\cite{Salam1,Salam2} to laser atomic traps experiments~\cite{Nature}. An immediate and important question that naturally arises in this fascinating field is: {\it How does the critical temperature of an asymmetric system depend on its asymmetry?} 

The critical temperature is the temperature at which the gap vanishes signing that the system is in the normal state with unpaired particles. If the mass and density asymmetries are large, the attraction is weak and we expect that in this asymmetrical conditions the critical temperature is smaller than that of the symmetric system. If the asymmetries increase more, even a small amount of heat is enough to break the Cooper pairs and disrupt superfluidity.

In this paper we investigate two species systems, and answer this question for two possibilities for the particle's masses and fixed chemical potentials asymmetries. We show that in a fermionic system constrained to $m_a \mu_a=m_b \mu_b$, with $m_a \neq m_b$ and $\mu_a \neq \mu_b$, the critical temperature is a slightly decreasing function of the asymmetry. We find a generalization for the expression relating the critical temperature and the zero temperature gap parameter.

\subsubsection{The Model}

As mentioned earlier, several candidates for pairing of unequal Fermi surfaces have been proposed in the literature. The Sarma phase~\cite{Sarma} and the prior version of the BP phase~\cite{Wilczek1}, are unstable~\cite{Heron1,Heron2,Wilczek3}, and will not be considered. Regarding the LOFF state, since this phase can exist only in a very narrow window of asymmetry for the Fermi surfaces, we do not consider this realization of superfluidity in the present calculations. The investigation of the critical temperature in the MP (with fixed particle densities)~\cite{Heron1,Heron2}, and in the clarified version of the BP phase\footnote{This version takes into account the momentum structure of the interaction and large mass asymmetry.} (with fixed chemical potentials)~\cite{Wilczek3}, must be considered and will be discussed elsewhere.

We employ the Bardeen-Cooper-Schrieffer (BCS) model to derive the Ginzburg-Landau (GL) theory, following the modern formulation developed by Sakita \cite{Sakita}. For the sake of completeness, we also derive the gap equation for an asymmetrical fermionic system at finite temperature by the variational method. The derivation is shown in the appendix, and the results agree, as it should. 

We consider an asymmetrical nonrelativistic dilute system of $a$ and $b$ fermion species\footnote{In (relativistic) quark matter, the study of the critical temperature from the point of view of the GL approach, has been carried out in Refs.~\cite{Iida1,Iida2}.}, having masses $m_a$ and $m_b$, with Fermi momentum $P_F^{j}=\sqrt{2m_j \mu_j}$, $j=a, b$. 

\begin{figure}[t]
\includegraphics[height=2in]{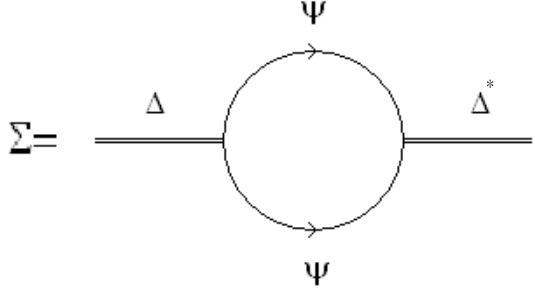}
\caption{\label{Mixed}\textit{The one-loop diagram contribution to $\Gamma[\Delta, \Delta^*]$ with two external lines. } }
\end{figure}

Let us begin with the partition function

\beq
\label{gf1}
Z=\int D[\psi_{a,b}] D[\psi_{a,b}^\dagger] e^{[- S(\psi, \psi^\dagger)]},
\eeq
where $S(\psi, \psi^\dagger)=-\int d \tau \int dx L$, with $L$ being the BCS Lagrangian

\beq
\label{BCS1}
L= \sum_{i=a,b} \psi_i^\dagger(x) \left(-\partial_{\tau} + \frac{\bigtriangledown^2}{2 m_i} + \mu_i \right) \psi_i(x) +
g \psi_{a}^\dagger(x) \psi_{b}^\dagger(x) \psi_{a}(x) \psi_{b}(x),
\eeq
where $g>0$. Introducing auxiliary fields via the Hubbard-Stratonovich transformation, we find an effective action, in which the mean field BCS Lagrangian is expressed by

\bea
\label{BCS2}
L_{MF}= \sum_{i=a,b} \psi_i^\dagger(x) \left(- \partial_{\tau} + \frac{\bigtriangledown^2}{2 m_i} + \mu_i \right) \psi_i(x)
 + \\
 \nonumber
 \Delta(x) \psi_{a}^\dagger(x) \psi_{b}^\dagger(x) + \Delta^*(x) \psi_{b}(x) \psi_{a}(x) + \frac{|\Delta(x)|^2}{g} .
\eea
Introducing a source for the $\Delta(x)$ field, we write
\beq
\label{gf3}
Z  = {\cal N} \int D[\psi] D[ \psi^\dagger] D[\Delta] D[\Delta^{\dagger}]~ e^{\int d \tau \int dx  \left[ L_{MF} +j^* \Delta(x) + j \Delta^*(x) \right]}
\eeq
The partition function is $Z=Z[j, j^*]_{j=j^*=0}$ and the generating functional for the connected Green's functions is defined as
\beq
\label{gf}
W[j, j^*]=\ln Z[j, j^*].
\eeq
The Legendre transformation of $W[j, j^*]$ is 
\beq
\label{g}
\Gamma[\Delta, \Delta^*]=\int d^4x (j^* \Delta(x) + j \Delta^*(x))- W[j, j^*].
\eeq
From all one-loop diagrams contributing to $\Gamma[\Delta, \Delta^*]$, we evaluate only the one which gives a contribution for the 
$\Delta^2$ term. It is shown in Fig.~(\ref{Mixed}).

\subsubsection{The one loop correction to the BCS gap parameter}

The effective action up to one-loop is
\beq
\label{ea}
\Gamma[\Delta,\Delta^*]=\alpha |\Delta|^2 +\beta |\Delta|^4  -c \Delta^* \frac{1}{8 M} {\vec{\nabla}}^2 \Delta,
\eeq
where $M=\frac{m_a m_b}{m_a + m_b}$ is the reduced mass, $\alpha=\frac{1}{g}-A$, and $A$ is the momentum independent contribution from $\Sigma(q_0=0,\vec{q})$, which is given by

\beq
\label{S}
\Sigma(q_0=0,\vec{q})=\sum_{n,k}\frac{1}{(i\omega_n-\omega^a(k))(-i\omega_n-\omega^b(k'))}=A+B {\vec{q}}^2 + \ldots,
\eeq
where $\omega_n=(2n+1)\pi T$, $\omega^a(k)=\frac{k^2}{2m_a}-\mu_a$, $\omega^b(k')=\frac{(\vec{k}+\vec{q})^2}{2 m_b}-\mu_b$. 

It is worth noting that Eq.~(\ref{ea}) is valid only in the vicinity of a second order phase transition, since it is based on the assumption that the gap parameter is small~\cite{Schafer}. While it is important that $\Delta$ be small so that higher order terms can be neglected, it is also important that the fermions have a finite gap so that they can be properly integrated out to obtain the effective potential. It is this latter condition that renders the present analysis appropriate only when the Fermi surfaces are not mismatched.

After the frequency summation, employing the imaginary time formalism of finite temperature field theory, we obtain

\bea
\label{f1}
A=\int \frac{d^3 k}{(2 \pi)^3} \frac{1}{\omega + \omega'} \left[1-\frac{1}{e^{\beta \omega}+1}- \frac{1}{e^{\beta \omega'}+1} \right],
\eea
where, for short, we have defined $\omega = \omega^a(k)$, $\omega' = \omega^b(k)=\frac{k^2}{2m_b}-\mu_b$ and $\beta=\frac{1}{k_B T}$, where $k_B$ is Boltzmann's constant that we will set equal to one. Changing the variable of integration from $k$ to $\epsilon$, we get
\bea
\label{f2}
A=\int_0^{\W_C} \rho(\epsilon) \frac{d \epsilon}{\epsilon} \left[1-\frac{1}{e^{\beta \omega(\epsilon)}+1}- \frac{1}{e^{\beta \omega'(\epsilon)}+1} \right]\\
\nonumber
\approx \rho(0)\int_0^{\W_C}  \frac{d \epsilon}{\epsilon} \left[1-\frac{1}{e^{\beta \omega(\epsilon)}+1}- \frac{1}{e^{\beta \omega'(\epsilon)}+1} \right],
\eea
where $\rho(0)=\frac{M k_F}{2 \pi^2}$ is defined as the density of states at the Fermi level, with $k_F=\sqrt{2 M \mu}$ being the ``average'' Fermi surface and $\mu = \mu_a + \mu_b$. We have also defined $\W_C(\Lambda)=\omega(\Lambda)+\omega'(\Lambda)$, where $\Lambda$ is the cutoff in the momentum integral. The energies are given by

\bea
\label{f22} 
\omega(\epsilon)=\frac{M}{m_a} \epsilon+ \mu\frac{M}{m_a}-\mu_a,\\
\nonumber 
\omega'(\epsilon)=\frac{M}{m_b} \epsilon+ \mu\frac{M}{m_b}-\mu_b.
\eea 
After simple algebra, Eq.~(\ref{f2}) can be written as

\beq
\label{f3}
A=\frac{\rho(0)}{2} \int_0^{\W_C}  \frac{d \epsilon}{\epsilon} \left[\tanh \left(\frac{\beta \omega(\epsilon)}{2} \right) +
\tanh \left(\frac{\beta \omega'(\epsilon)}{2} \right) \right].
\eeq
If we define
\begin{equation}
\label{eta}
\eta =\frac{M}{m_{a}}\mu -\mu _{a}=-\left( \frac{M}{m_{b}}\mu -\mu_{b}\right),
\end{equation}
then we can write Eq.~(\ref{f3}) with the aid of Eqs.~(\ref{f22}) and (\ref{eta}) as

\begin{eqnarray}
\label{eqmdif}
A &=&\frac{\rho (0)}{2}\left\{ \int_{0}^{\W_{C}}\frac{d\varepsilon }{%
\varepsilon }\tanh \left( \frac{\beta }{2}\frac{M}{m_{a}}\varepsilon +\frac{%
\beta }{2}\eta \right) +\int_{0}^{\W_{C}}\frac{d\varepsilon }{\varepsilon }%
\tanh \left( \frac{\beta }{2}\frac{M}{m_{b}}\varepsilon -\frac{\beta }{2}%
\eta \right) \right\} \\
 \nonumber
&=&\frac{\rho (0)}{2}\left\{ \int_{0}^{\lambda _{a}}\frac{dx}{x}\tanh \left(
x+a\right) +\int_{0}^{\lambda _{b}}\frac{dx}{x}\tanh \left( x-a\right)
\right\},
\end{eqnarray}
where we have defined
\begin{equation}
\lambda _{a}=\frac{\beta }{2}\frac{M}{m_{a}}\W_{C},\,\,\,\,\lambda _{b}=\frac{%
\beta }{2}\frac{M}{m_{b}}\W_{C}\text{ and }a=\frac{\beta }{2}\eta.
\end{equation}
If we take $m_{b} \geq m_{a}$, then $\lambda _{b} \leq \lambda _{a}$ and Eq.~(\ref{eqmdif}) can be written as

\begin{equation}
\label{eqLa}
A=\frac{\rho (0)}{2}\left\{ \int_{0}^{\lambda _{a}}\frac{dx}{x}\tanh \left(
x+a\right) +\int_{0}^{\lambda _{a}}\frac{dx}{x}\tanh \left( x-a\right)
-\int_{\lambda _{b}}^{\lambda _{a}}\frac{dx}{x}\tanh \left( x-a\right) \right\}.
\end{equation}
We can solve the first two integrals in the r.h.s. of Eq.~(\ref{eqLa}) employing the residue theorem and the last one can be easily solved if we observe that $\lambda _{a}$ and $\lambda _{b}$ correspond to the regions where $\tanh (x+a)\approx \tanh (x-a)\approx 1 $. Thus,

\begin{equation}
\label{eqf}
A=\frac{\rho (0)}{2}\left\{  \ln \left( \frac{\lambda_{a}^{2}}{\pi ^{2}}\right) -{\cal F}(a)  -\ln \left( \frac{\lambda _{a}}{\lambda _{b}}\right) \right\},
\end{equation}
where ${\cal F} (a)= \Psi (\frac{1}{2}+\frac{ia}{\pi })+\Psi (\frac{1}{2}-\frac{ia}{\pi })$ with $\Psi$ being the digamma function, defined as $\Psi(z)=\frac{\Gamma'(z)}{\Gamma(z)}$, where z is a complex number with a positive real component, $\Gamma$ is the gamma function, and $\Gamma'$ is the derivative of the gamma function. We also have that ${\cal F}(0)=-2\gamma -4\ln (2)$, where $\gamma$ is the Euler's constant.

The critical temperature is the solution of the equation 
\beq
\label{tc0}
\alpha= \frac{1}{g}-A=0.
\eeq
Then we write

\beq
\label{tc1}
\frac{1}{g \rho (0)} - \ln \left(\beta \sigma \frac{\omega_D}{\pi} \right)=-\frac{1}{2} {\cal F} (a),
\eeq
where $\sigma \equiv \frac{M}{\sqrt{m_a m_b}}$ is a dimensionless parameter reflecting the mass asymmetry, and we have used the fact that $\W_C=2\omega_D$.

The BCS gap in the weak coupling limit, $\rho(0)g<<1$, is given by $\Delta_0=2\omega_D~ e^{-1/\rho(0)g}$. Since the gap which minimizes the free-energy of the asymmetric system retains the same size as in the symmetric case ($\Delta_0$) until a value for the asymmetry where the pairing is not afforded any more~\cite{Paulo,Sedrakian,Toki,Heron1,Heron2}, we rewrite Eq.~(\ref{tc1}) as 

\beq
\label{tc2}
T_c=\frac{\sigma\Delta_0}{2 \pi} e^{-\frac{1}{2} {\cal F}(a_c)},
\eeq
where $a_c=\frac{\beta_c}{2} \eta = \frac{\beta_c}{2} \frac{m_b \mu_b- m_a \mu_a}{m_a + m_b}$. We evaluate Eq.~(\ref{tc2}) in the two possible configurations for the particles masses and chemical potentials constrained to $P^a_F=P^b_F$, which are the situations encountered when the fermions are fully gapped. In these cases we can obtain analytical solutions for the critical temperature, due to the simple form of the term ${\cal F} (a_c = 0)$, as showed below. 

In the investigation of the phase transition when the Fermi surfaces are mismatched, one needs to compare the thermodynamic potentials of the superfluid and normal states. In fact, as showed in~\cite{Paulo,Toki,Heron1,Heron2} by the behavior of the free-energy as a function of the gap for several asymmetries in the chemical potentials, an asymmetrical fermion system stays in the superfluid phase until a maximum value for the difference in chemical potentials is reached. After this maximum value, there is a first order phase transition to the normal phase.

\subsubsection{Equal Chemical Potentials and Masses}
\label{case1}

This case configures the symmetric system, whose critical temperature is recovered for $m_a=m_b,~\mu_a=\mu_b$, resulting 
${\cal F}(a_c=0)=-2 \ln(4 e^{\gamma})$, and $\sigma=\frac{1}{2}$, giving the well known BCS result

\beq
\label{tc3}
T_c=\frac{e^{\gamma}}{\pi} \Delta_0 \equiv T_c^{sym}.
\eeq
We notice that both the zero temperature gap parameter $\Delta_0$ and the symmetric critical temperature $T_c^{sym}$ depend on the product $\rho(0)g$, but not their ratio.

\subsubsection{Equal Fermi Surfaces with Mismatched Chemical Potentials and Masses}
\label{case2}

\begin{figure}[t]
\includegraphics[height=3in]{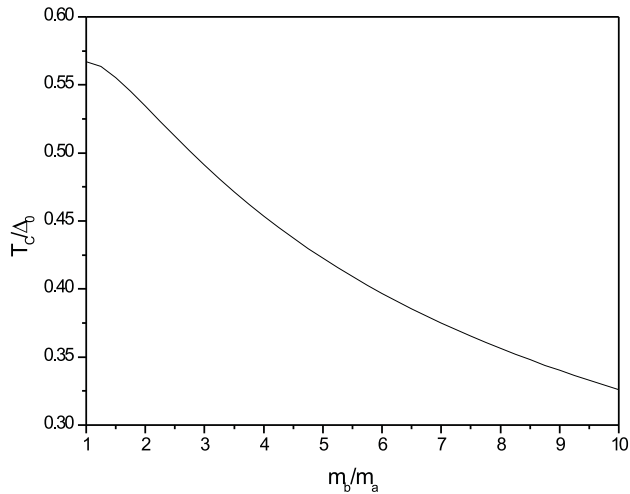}
\caption{\label{TempC1}\textit{$T_c / \Delta_0$ as a function of $m_b/m_a$ for a system constrained to $m_a \mu_a =m_b \mu_b$. } }
\end{figure}

This situation is also characterized by $P^a_F=P^b_F$, which implies $m_a \mu_a=m_b \mu_b$, however with $m_a \neq m_b$, $\mu_a \neq \mu_b$. This is achieved by setting $a_c =0$ in Eq.~(\ref{tc2}), yielding

\beq
\label{tc4}
T_c(P^a_F=P^b_F)= 2\sigma \frac{e^{\gamma}}{\pi} \Delta_0 = 2\sigma T_c^{sym}.
\eeq
We note that if we set $m_a=m_b$ in the equation above (which implies $\sigma=\frac{1}{2}$), then we have the symmetric Fermi gas, since we would also have $\mu_a=\mu_b$, due to the constraint $m_a \mu_a =m_b \mu_b$. We can observe from Eq.~(\ref{tc4}) that the critical temperature for the system constrained to $m_a \mu_a =m_b \mu_b$ goes with $2 \sqrt{\frac{m_a}{m_b}}~T_c^{sym}$ for $m_b$ greater than $m_a$ and approaches zero for $m_b>>m_a$. This shows that the pair formation is disfavored for very large mass asymmetry. The same conclusion has been found for the case of fixed number of particles~\cite{Heron3}. In Fig.~(\ref{TempC1}) we show the ratio $T_c / \Delta_0$ as a function of $m_b/m_a$. As one can see, $T_c / \Delta_0$ is a smooth function of the mass asymmetry, and goes to zero for $m_b>>m_a$. 

A mass ratio in the order of, or larger than, $m_b/m_a=50$, as used in Ref.~\cite{Wilczek3} would, obviously, have a smaller critical temperature than that of the symmetric gas. Although we did not consider in our calculations the momentum structure of the interaction employed in Ref.~\cite{Wilczek3}, we believe that our results would persist, at least qualitatively, after this consideration. 

\subsubsection{Conclusions}

We have calculated the critical temperature of an asymmetrical Fermi system in two configurations for the masses and chemical potentials of the two species that form Cooper pairs. Among those cases we have investigated, the one constrained to $m_a \mu_a = m_b \mu_b$, with $m_a \neq m_b$, $\mu_a \neq \mu_b$, is particularly interesting, for which we found a generalization for the expression relating the critical temperature and the zero temperature gap $\Delta_0$. Namely: $\frac{\Delta_0}{T_c}=\frac{\pi}{e^{\gamma}} \frac{1}{2\sigma} \approx 1.76 \frac{1}{2\sigma}$, where $\sigma = \frac{\sqrt{m_a m_b}}{m_a + m_b}=\frac{\sqrt{\mu_a \mu_b}}{\mu_a + \mu_b}$ and constitutes an {\it universal constant}, for given $m_a$ and $m_b$ (or $\mu_a$ and $\mu_b$), independent of $g$ and $\rho(0)$. Another remarkable feature of this result is its independence of any cutoff parameter. This is because Eq.~(\ref{tc2}) is quite insensitive to the regularization procedure.

We believe that the results achieved in this work could, in principle, be tested experimentally in, for example, experiments involving $^{6}Li$ or $^{40}K$ in atomic traps~\cite{DeMarco1,Ohara2,Truscott,DeMarco2,Ohara,Granade,Regal,Chin,Zwierlein,Kinast3,Bartenstein,Kinast4,Kinast,Kinast2}. Evidences of superfluidity in these systems were observed both microscopically, observing the pairing of fermionic atoms~\cite{Regal,Chin,Zwierlein} and macroscopically, due to anisotropic expansions~\cite{Ohara}, collective excitations~\cite{Kinast3,Bartenstein,Kinast4} and heat capacity measurements~\cite{Kinast}. In these experiments, the strength of the pairing interaction can be controlled by an applied magnetic field, for instance. Also, the density, the number of each species and the trapping potential can be altered. Thus, the species on the trap differ by theirs spin/pseudo-spin projections, and also by theirs densities, characterizing the asymmetry of the system.

Current experiments~\cite{Kinast,Kinast2,Kinast3,Kinast4} produce temperatures down to about $0.05T_{F}$, where $T_{F}$ is the Fermi temperature for a noninteracting gas with the same number of atoms and trap conditions as the experiment, typically of order of $\mu K$. However, a weakly interacting Fermi gas requires much lower $T$ to achieve superfluidity. For the conditions of these experiments, the mean field approximation with an interaction energy proportional to the scattering length is not valid. However, the mean field approximation with a unitary limit appears approximately valid, furnishing a good agreement with predictions of the collective frequencies, and a very good agreement on the transition temperature~\cite{Thomas}. Thus, even when the measurements are done in strongly interacting Fermi gases, mean field theory had qualitatively explained the behavior of these systems, and we expect that our weak coupling mean field BCS results should also be valid qualitatively. In particular, we believe that the numerical factor $1.76\frac{1}{2 \sigma}$ can be tested experimentally, provided the mass asymmetry is not so large.

When the Fermi surfaces are mismatched the phase transitions will be of first order and the present formalism is insufficient to determine the critical temperature. We will come back to this issue soon~\cite{Heron4}. For these cases $m_a \mu_a \neq m_b \mu_b$, and $\frac{\Delta_0}{T_c}$ will not be an universal number. This ``non universality'' is also manifested in the BP at finite temperature \cite{Liao} and in (dense) neutral quark matter \cite{Andreas,Igor,Kenji}, which is essentially asymmetrical \footnote{The up, down and strange quark Fermi surfaces, which are candidates to form {\it color superconductivity} are mismatched.}. Still in quark matter, in Ref. [39] was developed a systematic method of QCD expansion of the transition temperature, motivated by the non-BCS scaling of the gap parameter with coupling. In this work, the relation between the zero temperature energy gap and the critical temperature has a non BCS form too.

We find that large mass and chemical potentials asymmetries lower the critical temperatures substantially and compromise the stability of the system. This happens because any small thermal excitation breaks the (weakly bound due to the large asymmetry) pairs, and destroys superfluidity.

\begin{acknowledgments}
One of us, H.~Caldas, would like to thank P. Bedaque for valuable suggestions, Hai-Cang Ren for helpful discussions, and John Thomas for useful discussions on experimental issues related to Fermi systems. We thank the referee for pointing out to us a way to treat first order phase transitions. The authors acknowledge financial support by CNPq/Brazil.

\end{acknowledgments}

\subsubsection{Appendix: Solution by the Variational Method}

We now derive the gap equation at finite temperature for an asymmetrical fermion system, in order to determine the critical temperature. We follow the usual derivation of the textbooks, however extending the analysis for the asymmetrical systems we are investigating. Let us define $f_k$ as the probability of an $a$ particle with momentum $\bf{k}$ is excited, and similarly $g_k$ as the probability of a $b$ particle with momentum $\bf{-k}$ is excited. Then, the entropy for an asymmetrical fermion gas is found to be

\begin{equation}
\label{ap1}
S=-\sum_{k} \left\{ f_k \ln(f_k) + (1-f_k) \ln(1-f_k) + g_k \ln(g_k) + (1-g_k) \ln(1-g_k) \right\}.
\end{equation}
The free energy or thermodynamic potential is written as

\begin{equation}
\label{ap2}
F=E-TS,
\end{equation}
where $E$ is the internal energy

\bea
\label{ap3}
E=\sum_{k} \left\{ \epsilon_k^a [ (1- f_k -g_k)U_k^2+ f_k] + \epsilon_k^b [ (1- f_k -g_k)U_k^2+ g_k] \right\}\\
\nonumber 
-g \sum_{k,k'} U_{k'} V_{k'} U_k V_k (1-f_k - g_k)(1-f_{k'} - g_{k'}).
\eea
Here we have defined the particles energies relative to their Fermi surfaces in terms of our previous definitions $\epsilon_k^a \equiv \omega = \frac{k^2}{2 m_a}-\mu_a$ and $\epsilon_k^b \equiv \omega' = \frac{k^2}{2 m_b}-\mu_b$. From the minimizations

\bea
\label{ap4}
\frac{\delta F}{\delta f_k}=0,\\
\nonumber
\frac{\delta F}{\delta g_k}=0,\\
\nonumber
\frac{\delta F}{\delta U_k}=0,\\
\nonumber
\eea
we find, respectively,

\begin{equation}
\label{ap5}
f_k=1/(e^{\beta {\cal{E}}_k^a}+1),
\end{equation}

\begin{equation}
\label{ap6}
g_k=1/(e^{\beta {\cal{E}}_k^b}+1),
\end{equation}

\begin{equation}
\label{ap7}
U_k^2=\frac{1}{2} \left(1+ \frac{\epsilon_k^{+}}{E_k} \right),
\end{equation}
where ${\cal{E}}_k^{a,b}=\pm \epsilon_k^{-} + E_k$ are the quasiparticle excitations, with $E_k^2={\epsilon_k^{+}}^2+\Delta^2(T)$ and $\epsilon_k^{\pm} \equiv \frac {\epsilon_k^a \pm \epsilon_k^b}{2}$. In the definition of  ${\cal{E}}_k^{a,b}$ we have also defined

\begin{equation}
\label{ap8}
\Delta(T)=g \sum_{k} U_k V_k (1-f_k - g_k).
\end{equation}
Since $V_k^2=1-U_k^2$, then $U_k V_k=\frac{\Delta}{2 E_k}$, and the equation above can be written as

\begin{equation}
\label{ap9}
1=g \sum_{k} \frac{1}{2 E_k} \left( 1-f_k - g_k \right).
\end{equation}
At the critical temperature $\Delta=0$ in Eq.~(\ref{ap9}), and we have

\begin{equation}
\label{ap10}
1=g \sum_{k} \frac{1}{\epsilon_k^a+\epsilon_k^b} \left(1-\frac{1}{e^{\beta_c \epsilon_k^a}} - \frac{1}{e^{\beta_c \epsilon_k^b}} \right),
\end{equation}
which is equation~(\ref{tc0}). The solution for $T_c$ from (\ref{ap10}) follow as in the body of the paper.

Important remarks are now in order:\\ 

{\bf 1.} Although Eqs.~(\ref{ap9}) and~(\ref{ap10}) require regularization, the regulator dependence cancels from the result~(\ref{tc2}).\\

{\bf 2.} We have obtained an equivalence between two approximations to identify the zero temperature quasiparticle excitations. Introducing the temperature via the variational method, minimizing the thermodynamic potential with respect to the excitations probabilities in thermal equilibrium and then taking the zero temperature limit is equivalent to diagonalize the (zero temperature) mean field Hamiltonian (as done in Refs.~\cite{Heron1,Heron2}) and obtain the excitations energies of the fermion quasiparticles.\\

{\bf 3.} If the fermions are fully gapped, as is the case here, the gap parameter depends on the asymmetries only at finite temperature $(0<T<T_c)$. At zero temperature the excitations probabilities vanish, and the gap depends only on ``averages'', through $\epsilon_k^{+}$. When there are gapless excitations, the gap depends on the asymmetries even at $T=0$~\cite{Sarma,Wilczek2,Heron1,Heron2}.

\end{document}